\documentclass[aps,prd,preprint,tightenlines,showpacs,nofootinbib]{revtex4}

\usepackage{epsfig}
\usepackage{graphicx}

\newcommand{\tr}{ {\rm tr} }

\begin{document}
\title{Einstein-Yang-Mills-Lorentz black holes}

\author{Jose A. R. Cembranos\footnote{E-mail: cembra@fis.ucm.es},
and Jorge Gigante Valcarcel\footnote{E-mail: jorgegigante@ucm.es}
}

\affiliation{Departamento de  F\'{\i}sica Te\'orica I, Universidad Complutense de Madrid, E-28040 Madrid, Spain}


\pacs{04.70.Bw, 04.40.-b, 04.20.Jb, 11.10.Lm, 04.50.Kd, 04.50.-h}


%
%
%

\begin{abstract}
Different black hole solutions of the coupled Einstein-Yang-Mills equations have been well known for a long time. They have attracted much attention from mathematicians and physicists since
their discovery. In this work, we analyze black holes associated with the gauge Lorentz group. In particular, we study solutions which identify the gauge connection with the spin connection. This ansatz allows one to find exact solutions to the complete system of equations. By using this procedure, we show the equivalence between the Yang-Mills-Lorentz model in curved space-time and a particular set of extended gravitational theories.
\end{abstract}
\bigskip
\maketitle

\section{Introduction}

The dynamical interacting system of equations related to non-abelian gauge theories
defined on a curved space-time is known as Einstein-Yang-Mills (EYM) theory. Thus,
this theory describes the phenomenology of  Yang-Mills fields \cite{Yang-Mills}
interacting with the gravitational attraction, such as the electro-weak model or
the strong nuclear force associated with quantum chromodynamics.
EYM model constitutes a paradigmatic example of the non-linear interactions
related to gravitational phenomenology. Indeed, the evolution of a spherical symmetric
system obeying these equations can be very rich. Its dynamics is opposite to the one
predicted by other models, such as the ones provided by the Einstein-Maxwell (EM)
equations, whose static behaviour is enforced by a version of the Birkhoff theorem.

For instance, in the four-dimensional space-time, the EYM equations associated with the
gauge group $SU(2)$, support a discrete family of static self-gravitating solitonic solutions,
found by Bartnik and McKinnon \cite{BarMck}. There are also \emph{hairy} black hole (BH) solutions,
as was shown by Bizon \cite{Biz,Gal}. They are known as colored black holes and can be
labeled by the number of nodes of the exterior Yang-Mills field configuration.
Although the Yang-Mills fields do not vanish completely outside the horizon, these solutions
are characterized by the absence of a global charge.  This feature is opposite to the one predicted
by the standard BH uniqueness theorems related to the EM equations,
whose solutions can be classified solely with the values of the mass,
(electric and/or magnetic) charge and angular momentum evaluated at infinity.
In any case, the EYM model also supports the Reissner-Nordstr\"om BH as
an embedded abelian solution with global electric and/or magnetic charge \cite{Yas}. It is
also interesting to mention that there are a larger variety of solutions
associated with different generalizations of the EYM equations extended with
dilaton fields, higher curvature corrections, Higgs fields, merons or cosmological constants
(see \cite{Canfora,Volkov-Galt'sov} and the references therein).

In this work, we are interested in finding solutions of the EYM equations
associated with the Lorentz group as gauge group. The main motivation for considering
such a gauge symmetry is offered by the spin connection dynamics. This connection
is introduced for a consistent description of spinor fields defined on curved space-times.
Although general coordinate transformations do not have spinor representations \cite{Hehl},
they can be described by the representations associated with the Lorentz group. In addition,
they can be used to define alternative theories of gravity \cite{Sciama}.

We shall impose the requirement that the spin connection is dynamical and its evolution is determined by the
Yang-Mills action related to the $SO(1,n-1)$ symmetry, where $n$ is the number of dimensions
of the space-time. In order to complete the EYM equations, we shall assume that gravitation
is described by the metric of a Lorentzian manifold. We shall find vacuum analytic solutions
to the EYM system by choosing a particular $ansatz$, which will relate the spin connection
to the gauge connection. Therefore, this assumption provides additional gravitational degrees of freedom besides the ones given by the standard case, so that all the BH configurations found by this approach are not associated with an internal symmetry group and they do not carry any classical hair (i.e. they constitute a class of non-hairy BH solutions in a pure gravity model).

This work is organized in the following way. First, in Section II, we present basic features
of the EYM model. In Section III, we show the general results based on the Lorentz group
taking as a starting point the spin connection, which yields exact solutions to the EYM equations
in vacuum. The expressions of the field for the Schwarzschild-de Sitter metric in a four-dimensional
space-time are shown in Section IV, where we also remark some properties of particular the solutions
in higher dimensional space-times. Finally, we classify the Yang-Mills field configurations
through Carmeli method in Section V, and we present the conclusions obtained from our analysis in Section VI.

\section{EYM equations associated with the Lorentz group}

The dynamics of a non-abelian gauge theory defined on a four-dimensional Lorentzian manifold
is described by the following EYM action:
\begin{equation}\label{action}
S=-\frac{1}{16\pi}\int d^{4}x\sqrt{-g}\, R  + \alpha \int d^{4}x\sqrt{-g}\, \tr(F_{\mu \nu} F^{\mu \nu}) \,,
\end{equation}
where $A_\mu=A_\mu ^a\,T^a$, $[A_\mu, A_\nu]=if^{abc}A_\mu ^a\,A_\nu ^b\,T^c$, and
$F_{\mu \nu}=F_{\mu \nu}^a\,T^a$, $F_{\mu \nu}^a=\partial_{\mu}A_{\nu}^a-\partial_{\nu}A_{\mu}^a+f^{abc}A_{\mu}^b\,A_{\nu}^c$.
Unless otherwise specified, we will use Planck units throughout this work
($G=c=\hbar=1)$, the signature $(+,-,-,-)$ is used for the metric tensor, and Greek letters denote covariant indices,
whereas Latin letters stand for Lorentzian indices.
The above action is called pure EYM, since it is related to its simplest form, without any additional field or matter
content (see \cite{Volkov-Galt'sov} for more complex systems).

The EYM equations can be derived from this action by performing variations with respect to the gauge connection:
\begin{equation}\label{gaugeeq}
\left(D_{\mu}\,F^{\mu\nu}\right)^{a}=0\,,
\end{equation}
and the metric tensor:
\begin{equation}\label{einsteineq}
R_{\mu\nu}-\frac{R}{2}g_{\mu\nu}=8\pi T_{\mu\nu}\,,
\end{equation}
where the energy-momentum tensor associated with the Yang-Mills field configuration is given by:
\begin{equation}\label{energytensor}
T_{\mu\nu}=4\alpha\,\tr\,\left(F_{\mu\rho}F_{\nu}^{\,\,\,\rho}-\frac{1}{4}g_{\mu\nu}F_{\lambda\rho}F^{\lambda\rho}\right)\,.
\end{equation}
As we have commented, the first non-abelian solution with matter fields was found numerically by Bartnik and McKinnon for the
case of a four-dimensional space-time and a $SU (2)$ gauge group \cite{BarMck}. We are interested in solving
the above system of equations when the gauge symmetry is associated with the Lorentz group $SO (1,3)$. In this case,
we can write the gauge connection as $A_\mu=A^{a b}\,_{\mu}\,J_{a b}$, where
the generators of the gauge group $J_{ab}$, can be written in terms of the Dirac gamma matrices:
$J_{ab}=i[\gamma_a,\gamma_b]/8$. In such a case, it is straightforward to deduce the
commutative relations of the Lorentz generators:
\begin{equation}
\left[J_{ab},J_{cd}\right]=\frac{i}{2}\,\left(\eta_{ad}\,J_{bc}+\eta_{cb}\,J_{ad}-\eta_{db}\,J_{ac}-\eta_{ac}\,J_{bd}\right)\,.
\end{equation}

\section{EYM-LORENTZ $ANSATZ$}

The above set of equations constitutes a complicated system involving a large number of degrees of freedom,
which interact not only under the regular gravitational attraction but also under the non-abelian gauge interaction.
It is not simple to find its solutions. We propose the following $ansatz$, which identifies the gauge connection with the spin connection:
\begin{equation}\label{ansatz}
A^{a b}\,_{\mu}=e^{a}\,_{\lambda}\,e^{b \rho}\,\Gamma_{\,\,\,\rho \mu}^{\lambda}+e^{a}\,_{\lambda}\,\partial_{\mu}\,e^{b \lambda}\,,
\end{equation}
with $e^{a}\,_{\lambda}$ the tetrad field \cite{MTW,Wald}, that is defined through the
metric tensor $g_{\mu\nu}=e^{a}\,_{\mu}\,e^{b}\,_{\nu}\,\eta_{ab}$; and
$\Gamma_{\,\,\,\rho \mu}^{\lambda}$ is the affine connection of a semi-Riemannian manifold $V_4$.

By using the antisymmetric property of the gauge connection with respect to
the Lorentz indices: $A^{ab}\,_{\mu}=-\,A^{ba}\,_{\mu}$, we can write the field strength tensor
as
\begin{equation}
F^{a b}\,_{\mu \nu}=\partial_{\mu} A^{a b}\,_{\nu}-\partial_{\nu} A^{a b}\,_{\mu}+A^{a}\,_{c \mu}\,A^{c b}\,_{\nu}-A^{a}\,_{c \nu}\,A^{c b}\,_{\mu}\,.
\end{equation}
Then, by taking into account the orthogonal property of the tetrad field $e_{a}\,^{\lambda}\,e^{a}\,_{\rho}=\delta_{\rho}^{\lambda}$,
the field strength tensor takes the form \cite{Utiyama,Yepez}:
\begin{equation}
F^{a b}\,_{\mu \nu}=e^{a}\,_{\lambda}\,e^{b \rho}\,R^{\lambda}\,_{\rho \mu \nu}\,,
\end{equation}
where $R^{\lambda}\,_{\rho \mu \nu}$ are the components of the Riemann tensor.

Thus, $F_{\mu\nu}=e^{a}\,_{\lambda}\,e^{b \rho}\,R^{\lambda}\,_{\rho \mu \nu}\,J_{a b}$
represents a gauge curvature and we can express the pure EYM equations (\ref{gaugeeq}) and (\ref{einsteineq})
in terms of geometrical quantities. On the one hand, Eq. (\ref{gaugeeq}) can be written as:
\begin{equation}\label{ymecs}\left(D_{\mu}\,F^{\mu\nu}\right)^{ab}=e^{a}\,_{\lambda}\,e^{b}\,_{\rho}\,\nabla_{\mu}\,R^{\mu \nu \lambda \rho}=0\,,\end{equation}
whereas, on the other hand, the standard Einstein equation given by Eq. (\ref{einsteineq}) has the following gravitational correction to the Einstein tensor:
\begin{equation}\label{energyan}
T_{\mu \nu}=2\alpha\,\left(R^{\sigma\omega}\,_{\mu\rho}R_{\sigma\omega\nu}\,^{\rho}-\frac{1}{4}g_{\mu \nu}R_{\sigma \omega \lambda \rho}R^{\sigma \omega \lambda \rho}\right)\,,
\end{equation}
which replaces Eq. (\ref{energytensor}).

\section{SOLUTIONS OF THE EYM-LORENTZ $ANSATZ$}

The EYM-Lorentz $ansatz$ described above reduces the problem to a pure gravitational system and simplifies the search for particular solutions. According to the second Bianchi identity for a semi-Riemannian manifold, the components of the Riemann tensor satisfy:

\begin{equation}
\nabla_{[\mu}\,R_{\lambda \rho]}\,^{\sigma \nu}=0\,.
\end{equation}

By contracting this expression with the metric tensor:

\begin{equation}\nabla_{[\mu}\,R_{\lambda \rho]}\,^{\mu \nu}=0\,.\end{equation}

By using the symmetries of the Riemann tensor:

\begin{equation}\nabla_{\mu}\,R^{\mu \nu}\,_{\lambda \rho}+\nabla_{\rho}\,R_{\lambda}\,^{\nu}-\nabla_{\lambda}\,R_{\rho}\,^{\nu}=0\,,
\end{equation}
with $R_{\lambda}^{\,\,\,\,\,\nu}$ the components of the Ricci tensor. Then, taking into account (\ref{ymecs}), we finally obtain:

\begin{equation}\nabla_{[\lambda}R_{\rho]\nu}=0\,.\end{equation}

The integrability condition  $R_{[\mu \nu|\lambda|}\,^{\sigma}R_{\rho] \sigma}=0$ for this expression is known
to have as only solutions \cite{LsTr}:

\begin{equation}
\label{ec}R_{\mu \nu}=b\,g_{\mu \nu}\,,
\end{equation}
where $b$ is a constant.

First, we shall analyze the case of a space-time characterized by four
dimensions. In such a case, $T_{\mu\nu}$ is trace-free and the solutions are scalar-flat. From the expression of this tensor in terms of the Weyl and Ricci tensors, the Einstein equations are:

\begin{equation}
R_{\mu\nu}-16 \pi \alpha \, C_{\mu \lambda \nu \rho}R^{\lambda \rho}=0\,,
\end{equation}
where $C_{\mu \lambda \nu \rho}=R_{\mu \lambda \nu \rho}-\left(g_{\mu [\nu}R_{\rho] \lambda}-g_{\lambda [\nu}R_{\rho] \mu}\right)+R g_{\mu [\nu}g_{\rho] \lambda}/3\,.$

Therefore, by using (\ref{ec}) and the condition $C_{\mu \lambda \nu}\,^{\lambda}=0$, the only solutions are vacuum solutions defined by $R_{\mu\nu}=0$ \cite{Fairchild,Debney}. Hence, for empty space, $T_{\mu\nu}=0$ and all the equations are satisfied
for well-known solutions \cite{Yi-Cheng}, such as the Schwarzschild or Kerr metric. We can also add a cosmological constant in the Lagrangian and generalize
the standard solutions to de Sitter or anti-de Sitter asymptotic space-times,
depending on the sign of such a constant. Note that these solutions are
generally supported for a large variety of different
field models and gravitational theories \cite{Cembranos:2011sr}.

It is worthwhile to stress that these conclusions contrast with the ones given by other classical BH solutions in higher derivative gravity, where the approach assumes the requirement of the metric formalism and it leads to a different system of variational equations \cite{Stelle}. Indeed, whereas the gauge and the Palatini formalisms are found to be equivalent by requiring the presence of a metric-compatible connection \cite{Obukhov}, it is shown that the latter also implies the metric formalism but the opposite is not true for theories endowed with this type of higher order curvature corrections in the Lagrangian \cite{Borunda}. Then, it is expected that alternative vacuum solutions may also arise in the framework of the higher derivative gravity \cite{Lu}.

On the other hand, although the EYM theory typically involves gauging internal degrees of freedom associated with fields coupled to gravity, our solutions are also compatible with other gauge gravitational theories, such as Poincar\'{e} Gauge Gravity (PGG) \cite{Kibble,Hayashi,Blag-Hehl}. This theory is based on the Poincar\'{e} group, which is also known as the inhomogeneous Lorentz group. Within this model, the external degrees of freedom (rotations and translations) are gauged and the connection is defined by $A_{\mu}=e^{a}\,_{\mu}\,P_{a}+\left(e^{a}\,_{\lambda}\,e^{b \rho}\,\Gamma^{\lambda}\,_{\rho \mu}+e^{a}\,_{\lambda}\,\partial_{\mu}\,e^{b \lambda}\right)J_{a b}$, where $P_{a}$ are the generators of the translation group. The equations corresponding to the Lagrangian (\ref{action}) in PGG are the same than the previous system of equations \cite{Obukhov}. However, PGG is less constrained than a purely quadratic YM field strength.

Once the metric solution is fixed by the particular boundary conditions,
the EYM-Lorentz $ansatz$ defined by Eq. (\ref{ansatz}) determines the
solution of the Yang-Mills field configuration. In order to characterize such a
configuration, it is interesting
to establish the form of the electric $E_{\mu}=F_{\mu \nu}\,u{^\nu}$,
and magnetic field $B_{\mu}=*F_{\mu \nu}\,u{^\nu}$,
as measured by an observer moving with four-velocity $u{^\nu}$.
In particular, for the Schwarzschild-de Sitter solution, one may find the following electric and magnetic
{\it projections} of the Yang-Mills field strength tensor in the rest frame of reference \cite{Tseytlin}:
\begin{equation}
E_r=\frac{\frac{4M}{r^3}+\frac{2\Lambda}{3}}{\sqrt{1-\frac{2M}{r}-\frac{\Lambda}{3}\,r^2}}\,J_{01}\,,
\end{equation}
\begin{equation}
E_\theta=-\,2r\,\left(\frac{M}{r^3}-\frac{\Lambda}{3}\right)\,J_{02}\,,
\end{equation}
\begin{equation}
E_\phi=-\,2r\,\sin \theta\,\left(\frac{M}{r^3}-\frac{\Lambda}{3}\right)\,J_{03}\,,
\end{equation}
\begin{equation}
B_r=\frac{\frac{4M}{r^3}+\frac{2\Lambda}{3}}{\sqrt{1-\frac{2M}{r}-\frac{\Lambda}{3}\,r^2}}\,J_{23}\,,
\end{equation}
\begin{equation}
B_\theta=2r\,\left(\frac{M}{r^3}-\frac{\Lambda}{3}\right)\,J_{13}\,,
\end{equation}
\begin{equation}
B_\phi=-\,2r\,\sin \theta\,\left(\frac{M}{r^3}-\frac{\Lambda}{3}\right)\,J_{12}\,.
\end{equation}

It is straightforward to check that the above solution verifies:
\begin{equation}
\tr\left(\vec{E}^2+\vec{B}^2\right)=0\,,
\end{equation}
and
\begin{equation}
\tr\left(\vec{E} \cdot \vec{B}\right)=0\,.
\end{equation}

It is also interesting to remark that
the family of solutions  provided by the EYM-Lorentz $ansatz$ is not restricted to the signature $(+,-,-,-)$.
It is also valid for the Euclidean case $(+,+,+,+)$. For the latter signature, the corresponding gauge group is $SO\,(4)$  and
the associated generators satisfy the following commutation relations:
\begin{equation}
[J_{ab},J_{cd} ]=\frac{i}{2}\,\left(\delta_{ad}\,J_{bc}+\delta_{cb}\,J_{ad}-\delta_{db}\,J_{ac}-\delta_{ac}\,J_{bd}\right)\,.
\end{equation}

The above solutions can also be generalized to a space-time with an arbitrarily higher number of dimensions.
For the n-dimensional case, the assumption of the $ansatz$ (\ref{ansatz})
in the EYM equations (\ref{gaugeeq}), (\ref{einsteineq}) and (\ref{energytensor}) is equivalent
to work with the following gravitational action in the Palatini formalism:
\begin{equation}\label{actionn}
S=\int d^{n}x\sqrt{-g}\,
\left\{- \frac{1}{16 \pi}R
+ 2^{\tilde{n}/2-3} \alpha R_{\lambda \rho \mu \nu}R^{\lambda \rho \mu \nu}
\right\}
\,,
\end{equation}
where $\tilde{n}=n$ and $\tilde{n}=n-1$ for even and odd $n$.

In such a case, the quadratic Yang-Mills correction takes the form of the one associated with a cosmological constant, in a similar
way to certain solutions of modified gravity theories, as the Boulware-Deser solution in Gauss-Bonnet gravity \cite{Boulware-Deser}.
For instance, for a de Sitter geometry, the Riemann curvature tensor is given by
\begin{equation}
R_{\lambda \rho \mu \nu} = \frac{2\Lambda}{(n-2)(n-3)} \,\left(g_{\lambda \mu}\,g_{\rho \nu}-g_{\lambda \nu}\,g_{\rho \mu}\right)\,.
\end{equation}
In this case, the geometrical correction associated with the Yang-Mills
configuration given by Eq. (\ref{energyan}) takes the form
\begin{equation}
T_{\mu \nu} = - \, 2^{\tilde{n}/2} \alpha\,\Lambda^2\frac{\left(n-1\right)\left(n-4\right)}{\left(n-2\right)^2\left(n-3\right)^2}\,g_{\mu \nu}\,.
\end{equation}

Therefore, $T_{\mu \nu}=0$ is a particular result associated with the four-dimensional space-time.

On the other hand, the equivalence between the Yang-Mills-Lorentz model in curved space-time
and a pure gravitational theory is not restricted to Einstein gravity. For example,
in the five-dimensional case, we can study the gravitational
model defined by the following action in the Palatini formalism:
\begin{equation}\label{action5}
S_{G}=\int d^{5}x\sqrt{-g}\,
\left\{\alpha_{0}+\alpha_{1}R
+\alpha_{2}R^2-4\alpha_{3}R_{\mu \nu}R^{\mu \nu}+\alpha_{4}R_{\lambda \rho \mu \nu}R^{\lambda \rho \mu \nu}
\right\}
\,.
\end{equation}

The above expression includes not only the cosmological constant (proportional to $\alpha_{0}$) and the
Einstein-Hilbert term (proportional to $\alpha_{1}$), but also quadratic contributions of the
curvature tensor (proportional to $\alpha_{2}$, $\alpha_{3}$ and $\alpha_{4}$). In this case, the addition
of the Yang-Mills action under the restriction of the Lorentz $ansatz$ (\ref{ansatz}) is equivalent to
work with the same gravitational model given by Eq. (\ref{action5}) with the
following redefinition of $\alpha_{4}$:
\begin{equation}
\alpha^{YM}_{4}=\alpha_{4}+\frac{\alpha}{2}
\,.
\end{equation}
It is particularly interesting to consider the model with $\alpha_{2}=\alpha_{3}=\alpha^{YM}_{4}$.
In such a case, the higher order contribution in the equivalent gravitational system is proportional to the
Gauss-Bonnet term. As is well known, this latter term reduces to a topological surface
contribution for $n=4$, but it is dynamical for $n\geq 5$. In particular, according to the Boulware-Deser solution,
the metric associated with the corresponding equations takes the simple form:
\begin{equation}
ds^2=A^2(r)\,dt^2-\frac{dr^2}{A^2(r)}-r^2d\Omega^{2}_{3}\,,
\end{equation}
where $d\Omega^{2}_{3}$ is the metric of a unitary three-sphere, and $A^2(r)$ is given by:
\begin{equation}
A^2(r)=1+\frac{r^2}{4\Upsilon}+\sigma\frac{r^2}{4\Upsilon}\sqrt{1+\frac{16\Upsilon M}{r^4}+\frac{4\Upsilon\Lambda}{3}}\,,
\end{equation}
with $\alpha_{0}/\alpha_{1}=-2\Lambda$, $\alpha_{2}/\alpha_{1}=\Upsilon$, and $\sigma=1$ or $\sigma=-1$.
Therefore, from the EYM point of view, the Yang-Mills field contribution modifies the metric solution in a very non-trivial way.
We can study the limit $\Upsilon \rightarrow 0$ in the Boulware-Deser metric. It is interesting to note that it does not
necessarily mean a weak coupling regime of the EYM interaction, since $\alpha^{YM}_{4} \rightarrow 0$ does not imply $\alpha \rightarrow 0$. It is convenient to distinguish between the branch $\sigma=-1$ and $\sigma=1$.
The first choice recovers the Schwarzschild-de Sitter solution for $\Upsilon=0$:
\begin{equation}
A^2_{\sigma=-1}(r) \simeq 1
-\frac{2M}{r^2}\left(1-\frac{2\Lambda\Upsilon}{3}\right)
-\frac{\Lambda}{6}\left(1-\frac{\Lambda\Upsilon}{3}\right)\,r^2
+\frac{8M^2\Upsilon}{r^6}
\,.
\end{equation}

When this metric is deduced from the equations corresponding to a pure gravitational theory,
the new contributions from finite values of $\Upsilon$ are usually interpreted as short distance corrections
of high-curvature terms in the Einstein-Hilbert action. From the EYM model point of view, these corrections originate with the Yang-Mills contribution interacting with the gravitational attraction.

On the other hand, the metric solution takes the following form in the EYM weak coupling limit
for the value $\sigma=1$:
\begin{equation}
A^2_{\sigma=1}(r) \simeq 1
+\frac{2M}{r^2}\left(1-\frac{2\Lambda\Upsilon}{3}\right)
+\frac{\Lambda}{6}\left(1+\frac{3}{\Lambda\Upsilon}-\frac{\Lambda\Upsilon}{3}\right)\,r^2
-\frac{8M^2\Upsilon}{r^6}\,.
\end{equation}

The corresponding geometry does not recover the Schwarzschild-de Sitter limit
when $\Upsilon \rightarrow 0$, and it shows ghost instabilities.

\section{CARMELI CLASSIFICATION OF THE YANG-MILLS FIELD CONFIGURATIONS}

In the same way that the Petrov classification of the gravitational field describes the possible algebraic symmetries of the Weyl tensor through the problem of finding their eigenvalues and eigenbivectors \cite{Petrov}, the Carmeli classification analyzes the symmetries of Yang-Mills fields configurations \cite{Carmeli}.

Let $\xi_{ABCD}$ be the gauge invariant spinor defined by  $\xi_{ABCD}=\frac{1}{4}\epsilon^{\,\dot{E}\dot{F}}\epsilon^{\,\dot{G}\dot{H}}\tr\left(f_{A\dot{E}B\dot{F}}\,f_{C\dot{G}D\dot{H}}\right)$,
with $f_{A\dot{B}C\dot{D}}=\tau_{A\dot{B}}^{\mu}\,\tau_{C\dot{D}}^{\nu}\,F_{\mu\nu}$
the spinor equivalent to the Yang-Mills strength field tensor written in terms of the generalizations of the unit and Pauli matrices,
which establish the correspondence between spinors and tensors. Let $\phi_{AB}$ be a symmetrical spinor.
Then, by studying the eigenspinor equation $\xi_{AB}\,^{CD}\,\phi_{CD}=\lambda\,\phi_{AB}$,
we can classify Yang-Mills field configurations in a systematic way.

This analysis can be applied to any of the EYM-Lorentz solutions but, for simplicity, we will illustrate the computation for the
EYM solution related to the Schwarzschild metric in four dimensions. We find the following invariants of the Yang-Mills field:
%
%
%
%
%
\begin{equation}
P=\xi_{AB}\,^{AB}=\frac{3M^2}{4r^6}\,,
\end{equation}
\begin{equation}
G=\eta_{ABCD}\,\eta^{ABCD}=\frac{3M^4}{32r^{12}}\,,
\end{equation}
\begin{equation}
H=\eta_{AB}\,^{CD}\,\eta_{CD}\,^{EF}\,\eta_{EF}\,^{AB}=\frac{3M^6}{256r^{18}}\,,
\end{equation}
\begin{equation}
S=\xi_{ABCD}\,\xi^{ABCD}=\frac{9M^4}{32r^{12}}\,,
\end{equation}
\begin{equation}
F=\xi_{AB}\,^{CD}\,\xi_{CD}\,^{EF}\,\xi_{EF}\,^{AB}=\frac{33M^6}{256r^{18}}\,,
\end{equation}
where $\eta_{ABCD}$ is the totally symmetric spinor $\xi_{\left(ABCD\right)}$, and $\xi_{ABCD}$ satisfies the equalities $\xi_{ABCD}=\xi_{BACD}=\xi_{ABDC}=\xi_{CDAB}$. Then, the characteristic polynomial
$p(\lambda')=\lambda'^3-G\lambda'/2-H/3$ associated with eigenspinor equation of $\eta_{ABCD}$
provides directly the eigenvalues of the corresponding $\xi_{ABCD}$. By taking
$\lambda=\lambda'+P/3$, we obtain the following results:
\begin{equation}
\lambda_1=\frac{M^2}{2r^6}\,,
\end{equation}
\begin{equation}
\lambda_{2,3}=\frac{M^2}{8r^6}\,.
\end{equation}

Thus, there are two different eigenvalues:  the first one is simple, whereas the second one is double.
There are three distinct eigenspinors and the corresponding Yang-Mills field is of type $D_{P}$, which is associated with
the Yang-Mills configurations of isolated massive objects.

\section{CONCLUSIONS}

In this work, we have studied the EYM theory associated with a $SO(1,n-1)$ gauge symmetry, where $n$ is the number
of dimensions associated with the space-time. In particular, we have derived analytical expressions for a large
variety of BH solutions. For this analysis, we have used an $ansatz$ that identifies the
gauge connection with the spin connection. We have shown that this $ansatz$ allows one to interpret
different known metric solutions corresponding to pure gravitational systems, in terms of equivalent EYM models.
We have demonstrated that this analytical method can also be applied successfully to the study of fundamental BH configurations. Such configurations usually differ from the given by the standard case, so that they are useful to improve the understanding of the resulting approach by showing the similarities and differences with respect to the present in other quadratic gravity theories (see \cite{Cap-Lau} and the references therein for a recent overview and additional BH solutions).

For the analysis of the corresponding Yang-Mills model with Lorentz gauge symmetry in curved space-time,
we have used the appropriate procedure in order to solve the equivalent gravitational equations,
which governs the dynamics of pure gravitational systems associated with the proper gravitational theory.
In particular, we have derived the solutions for the Schwarzschild-de Sitter geometry in a four-dimensional space-time
and for the Boulware-Deser metric in the five-dimensional case. For these solutions, we have specified the
corresponding pure gravitational theories. The algebraic symmetries associated with the Yang-Mills configuration related
to a given solution can be classified by following the Carmeli method. We have explicitly shown the equivalence
with the Petrov classification for the Schwarzschild metric in four dimensions.

In addition, numerical results obtained for these gravitational systems can
be extrapolated to the EYM-Lorentz model by following our prescription.
Through the gravitational analogy, one can also deduce the stability properties
of the EYM solutions or the gravitational collapse associated with such a system.
Here, we have limited the EYM-Lorentz $ansatz$ to the analysis
of spherical and static BH configurations, but it can be used to study
other types of solutions. For example, by using the same $ansatz$,
gravitational plane waves in modified theories of gravity may be interpreted as
EYM-Lorentz waves. We consider that all these ideas deserve further investigation
in future work.

\bigskip
\bigskip
\noindent
{\bf ACKNOWLEDGMENTS}

\bigskip
We would like to thank Luis J. Garay and Antonio L. Maroto for helpful discussions.
J.G.V. would like to thank Francisco J. Chinea for his useful advice.
This work has been supported by the MICINN (Spain) project numbers FIS2011-23000, FPA2011-27853-01,
Consolider-Ingenio MULTIDARK CSD2009-00064.
J.A.R.C. thanks
the SLAC National Accelerator Laboratory, Stanford (California, USA) and the University of Colima
(Colima, Mexico) for their hospitality during the latest stages of the preparation of this manuscript, and
the support of the {\it Becas Complutense del Amo} program and
the {\it UCM Convenio 2014} program for professors.

\end{document}